\documentclass{vgtc}                          

\graphicspath{{figures/}{pictures/}{images/}{./}} 

\usepackage{times}                     

\usepackage{tabu}                      
\usepackage{booktabs}                  
\usepackage{lipsum}                    
\usepackage{mwe}                       

\usepackage{mathptmx}                  
\usepackage{siunitx}
\usepackage{enumitem}
\usepackage{tabularx}

\usepackage{graphicx}
\usepackage{subcaption}
\usepackage{float}
\usepackage{multirow}
\usepackage{arydshln}
\usepackage{graphicx}
\usepackage{siunitx}
\usepackage{longtable}
\usepackage{ragged2e}
\usepackage{gensymb}
\usepackage{accsupp}

\usepackage[acronym,shortcuts]{glossaries}
\newacronym{BLV}{BLV}{blind or low vision}
\newacronym{DIY}{DIY}{do-it-yourself}
\newacronym{HCI}{HCI}{human-computer interaction}
\newacronym{HMD}{HMD}{head-mounted display}
\newacronym{IMU}{IMU}{inertial measurement unit}
\newacronym{LLM}{LLM}{large language model}
\newacronym{OandM}{O\&M}{orientation and mobility}
\newacronym{PSNR}{PSNR}{peak signal-to-noise ratio}
\newacronym{SLAM}{SLAM}{Simultaneous Localization and Mapping}
\newacronym{NeRF}{NeRF}{Neural Radiance Fields}
\newacronym{AR}{AR}{Augmented Reality}
\newacronym{VR}{VR}{Virtual Reality}
\newacronym{XR}{XR}{Extended Reality}
\newacronym{FOV}{FOV}{field of view}
\newacronym{ToF}{ToF}{Time of Flight}

\usepackage{xcolor}

\onlineid{0}

\vgtccategory{Research}

\vgtcinsertpkg

\title{Navigating the Last Mile: Evaluating Head- and Cane-Mounted Cameras for Egocentric Spatial Awareness}

\author{Apurv Varshney\thanks{e-mail: apurv@ucsb.edu (Corresponding Author)}\\ %
        \scriptsize University of California, Santa Barbara %
\and Lucas Nadolskis\thanks{e-mail: lgilnadolskis@ucsb.edu}\\ %
     \scriptsize University of California, Santa Barbara %
\and Tobias Höllerer\thanks{e-mail: holl@cs.ucsb.edu}\\ %
     \scriptsize University of California, Santa Barbara %
\and Michael Beyeler\thanks{e-mail: mbeyeler@ucsb.edu}\\ %
     \scriptsize University of California, Santa Barbara}

\teaser{
  \centering
  \includegraphics[width=.9\textwidth]{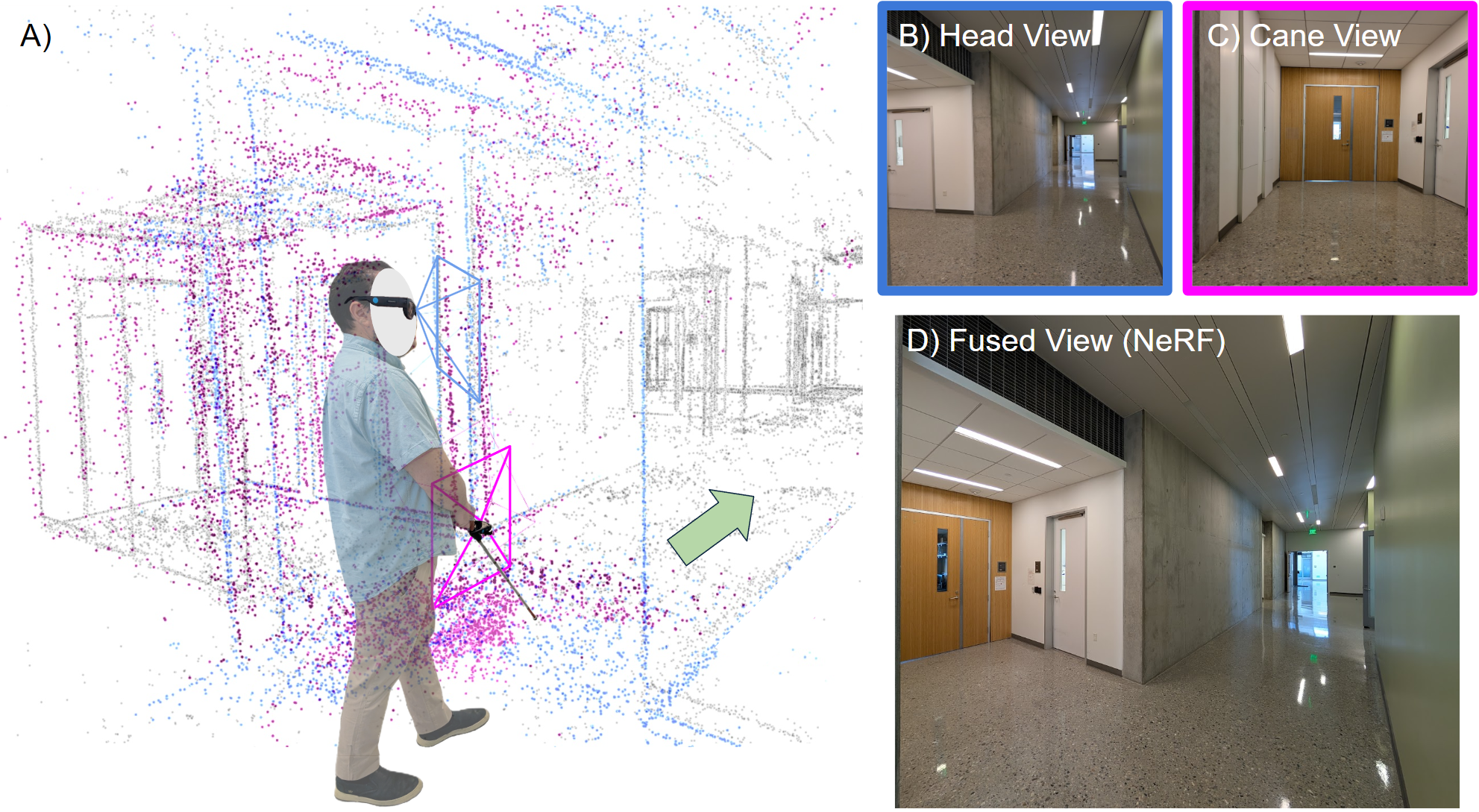}
  \caption{Evaluating head- vs.\ cane-mounted camera placements for assistive XR navigation in a complex ``last-mile'' indoor environment.
A) \Acf{SLAM} point cloud reconstructions demonstrate how head-mounted (blue) and cane-mounted (pink) camera vantages capture complementary spatial geometry (gray points denote unmapped regions).
B) Head-mounted view: forward-facing and gaze-aligned, providing the visual stability necessary for robust localization.
C) Cane-mounted view: dynamic, ground-level perspective that captures critical low-lying geometry beyond the cane's tactile reach.
D) Hybrid (head+cane) multi-view reconstruction via \acf{NeRF}: integrating both vantages synthesizes a high-fidelity 3D scene representation, combining tracking stability with dense geometric coverage to support downstream spatial reasoning.}
  \label{fig:teaser}
}

\abstract{Robust navigational guidance is an important XR application for both sighted and non-sighted populations. In this paper, we mainly focus on blind pedestrians, who continue to face ``last-mile'' challenges such as locating entrances and navigating cluttered spaces. While smartglasses and wearables are maturing, a foundational design question remains underexplored: where on the body should cameras be placed to best support navigation? We present a mixed-methods investigation that focuses on the question of camera placement for generating spatial data supporting ego-centric navigation. A survey of 10 blind cane users surfaced practices for last-mile navigation and perceptions of body-mounted XR devices. A controlled case study with a blind co-author compared head- and cane-mounted cameras using synchronized Project Aria glasses while traversing five real-world environments. Using Simultaneous Localization and Mapping (SLAM) and Neural Radiance Fields (NeRFs) as diagnostic probes, we find the central trade-off appears to be not simply head versus cane but localization stability versus near-ground coverage, with the combined head+cane view providing complementary information for scene reconstruction. We translate these findings into architectural considerations for hybrid XR systems that extend the cane without interfering with tactile and auditory cues.
} 


\begin{document}



\firstsection{Introduction}
\maketitle

Independent travel often breaks down for travelers at the ``last mile'': the final few meters of a journey, such as locating a doorway in a plaza or identifying the right bus at a stop (also known as the ``last 10-meters/yards'' problem~\cite{lastMeter2019,gleason_footnotes_2018,lock_portable_2017}). These moments demand precise spatial awareness and robust environment tracking beyond the physical reach of the long cane or guide dog, remaining among the most challenging operational environments for assistive \ac{XR} systems~\cite{unfamiliarindoorNeeds2022,streetnav2024,outdoorSurvey2023,gupta_towards_2020,kacorri_environmental_2018}.

Recent advances in wearable sensing have brought camera-equipped smartglasses and compact body-worn devices into everyday use~\cite{Envision2023SmartGlasses,projectAria}. 
As this hardware matures, a foundational system design question remains underexplored: \emph{where on the body should cameras be placed} to best support navigation for blind cane users? 
Prior work has explored head-, chest-, and cane-mounted configurations with distinct trade-offs: head-mounted cameras align with gaze and support landmarking~\cite{arthur_effects_2000,lee_accessing_2021} but can miss low-lying hazards; chest or belt placements offer stability and discretion~\cite{mathis_lifeinsight_2025} yet often duplicate head-level coverage; and cane-mounted sensors couple to tactile exploration~\cite{caneLessons2020} but their sweeping motion can blur images and disrupt pose estimation~\cite{slam2024}. Despite decades of assistive system proposals, camera placement itself has rarely been isolated as an independent design variable evaluated under controlled spatial computing conditions.

This paper treats camera placement as a first-class variable and examines its technical and experiential consequences for assistive XR navigation pipelines. 
We adopt a convergent mixed-methods design~\cite{creswell_designing_2011}, collecting qualitative (formative survey) and quantitative (controlled vantage study) data in parallel and integrating results at the interpretation stage.
We establish system requirements through an initial formative survey ($n=10$) probing blind cane users' last-mile operational strategies, spatial information needs, and form-factor constraints for emerging wearables~\cite{2019navigationsurvey,khan_technology-assisted_2018,deviceNeeds2023,herskovitz_hacking_2023}. We then conduct a controlled vantage study with an experienced blind co-author, directly comparing synchronized head- and cane-mounted Project Aria devices~\cite{projectAria} across five complex indoor and outdoor locations.

Both studies were shaped through an iterative co-design process with the blind co-author~\cite{codesign2019,mack2021we,brule2020review,codesignJingyi2019}, an experienced independent cane user and long-time tester of assistive technologies. 
The co-design partner collaborated on survey question design, helped interpret and thematically code open-ended responses, and co-selected sites and landmarks for the vantage study to reflect authentic \ac{OandM} challenges. They were the sole participant in the data collection for the controlled Camera Vantage Study presented in Section \ref{sec:vantage}, but had only limited involvement in the analysis of the collected data (helping with outcome-independent questions regarding the interpretation). 
This co-design approach ensured that our technical evaluations foregrounded authentic user priorities and practical form-factor limitations, rather than relying solely on isolated bench-testing, while also shielding against preferential bias. 

We operationalize the effects of vantage using two core spatial computing pipelines widely used in \ac{XR} navigation: \ac{SLAM} as a proxy for maintaining a stable pose estimate and sparse geometric map, and \ac{NeRF} as a proxy for generating high-fidelity, view-consistent 3D scene reconstructions suitable for downstream semantic reasoning.
While coarse positioning (e.g., GPS) suffices for block-level guidance, last-mile decisions demand doorway-scale precision and robust indoor tracking~\cite{turkstra_assistive_2025,kacorri_environmental_2018}, requiring \ac{XR} systems that can build and update persistent maps in situ.

Our analysis surfaces a core trade-off: head-mounted views provide the stability necessary for reliable localization, cane-mounted views capture complementary ground-level geometry useful for obstacle avoidance, and fusing both perspectives yields improved scene reconstruction quality. A stationary-cane control further suggests that motion, rather than the cane viewpoint itself, is the primary contributor to the reduced \ac{SLAM} metrics. 
In short, our contributions are threefold:
\begin{itemize}[topsep=0pt]
    \item \textbf{Community-grounded system constraints:} a survey that surfaces operational barriers, spatial information needs, and form-factor requirements for assistive \ac{XR} wearables.
    \item \textbf{Empirical characterization of vantage trade-offs:} a synchronized head-cane comparison across multiple environments, including a stationary-cane control study that isolates the impact of sweeping dynamics on computer vision pipelines.
    \item \textbf{Architectural considerations for hybrid or adaptive systems:} recommendations on multi-view integration and hybrid vantage strategies to balance localization stability with ground-level mapping for last-mile navigation.
\end{itemize}

\section{BACKGROUND}

We discuss related work on Navigation Challenges for Blind Travelers, Wearable XR Navigation Systems, and Emerging Spatial Pipelines before identifying Knowledge Gaps and reiterating our three main research questions.  

\subsection{Navigation Challenges for Blind Travelers}

Navigating unfamiliar or complex environments remains persistently difficult for blind cane users, who rely primarily on tactile and auditory rather than visual strategies~\cite{2019navigationsurvey,unfamiliarindoorNeeds2022,spatialLandmarks2023,lastMeter2019,supportExploration2023}. 
A long-running debate in assistive technology is whether systems should prioritize immediate \emph{obstacle avoidance} or broader \emph{spatial awareness and destination finding}. 
While early technical systems emphasized local obstacle detection~\cite{caneLessons2020,dlbaseddevice2019}, \ac{AR} and \ac{XR} are increasingly positioned as critical accessibility tools capable of delivering rich, global spatial context~\cite{AR4VI}.

Environmental complexity poses significant challenges for these spatial computing pipelines. Indoors, wide-open layouts lack tactile anchors, prompting approaches that rely on pre-mapped AR environments\cite{2021ruvolo} or crowdsourced visual route assistance~\cite{crowdsource2022}. Outdoors, traditional GPS lacks the doorway-scale precision required for the ``last mile''~\cite{streetnav2024}. To address this, modern \ac{XR} interventions are exploring richer forms of support, including AR-driven tele-guidance~\cite{2023guidance}, robotic companions~\cite{kuribayashi_wanderguide_2025,kamikubo_beyond_2025,kayukawa2023enhancing}, and conversational AI~\cite{kaniwayuka_chitchatguide_2024}. However, delivering this level of context-aware assistance requires highly robust computer vision localization, which fundamentally depends on where the tracking cameras are placed on the user's body.
Our focus here is to \emph{augment} proven tools (particularly the long cane) with computer vision capabilities rather than \emph{replace} them.

\subsection{Wearable XR Navigation Systems}

A wide spectrum of wearable AR aids has been proposed, spanning head-, chest-, and cane-mounted placements. Each offers distinct camera vantages and affordances, yet few have been systematically compared regarding their impact on downstream computer vision tasks.

\emph{Head-mounted systems} dominate both research and commercial accessibility applications, as evidenced by recent scoping reviews~\cite{2025review_hmd}. Devices like Envision Glasses~\cite{Envision2023SmartGlasses,OrCam_MyEye_3_Pro} and research platforms like Project Aria~\cite{projectAria} utilize head-worn cameras for landmark detection, text recognition, and SLAM mapping. Head-mounted views inherently align with user orientation and gaze~\cite{arthur_effects_2000,lee_accessing_2021}, providing the stable viewpoints necessary for robust visual-inertial odometry. However, critics note their limited field of view for low-lying hazards and immediate ground geometry~\cite{dlbaseddevice2019}. This trade-off between stable, gaze-aligned coverage and missed ground features is unresolved.

\emph{Chest- and belt-mounted systems} offer alternatives, providing stable camera rigs without head-weight~\cite{navbelt2003,BipedAI}. While advocates highlight tracking stability and reduced conspicuousness~\cite{wearabledevice2017}, these placements often duplicate head-level coverage without significantly improving ground-plane reconstruction~\cite{kubota_snapnav_2024}.

\emph{Cane-mounted systems} embed vision or depth sensors directly into the mobility cane, promising a tight coupling with tactile exploration and unparalleled coverage of the ground plane~\cite{stanfordcane2021,gonsher_smart_2023,zhao2018enabling,agrawal2022novel,Wewalk}. However, this vantage point introduces severe technical challenges. The dynamic, sweeping motion of the cane causes motion blur and rapid viewpoint shifts that frequently break visual SLAM pose estimation~\cite{slam2024,motioBlur2021,bamdad2025deep}. While modern collaborative and visual-inertial SLAM architectures can handle complex, multi-agent tracking~\cite{covins2021}, the aggressive local motion of a cane often exceeds standard algorithmic tolerances, limiting its use as an independent tracking node.

\emph{Multimodal XR Feedback} is the crucial counterpart to robust tracking. Once an XR system successfully maps an environment, it must convey spatial geometry without masking auditory cues. Recent work heavily explores multimodal and haptic AR interfaces to deliver this continuous spatial guidance, utilizing targeted audio~\cite{ARCore, 2022multimodal} alongside vibrotactile feedback~\cite{2025comparefeedback} to communicate complex spatial information safely.

Taken together, the literature shows that head-mounted systems are stable but incomplete, chest-mounted ones offer redundancy with modest ergonomic gains, and cane-mounted ones provide unique but noisy perspectives. 
Crucially, prior work typically fixes a single head-, chest-, or cane-mounted device as one component of an end-to-end prototype and evaluates that system as a whole, so camera placement is confounded with the rest of the design — sensor model, algorithm, and interface all vary together. Few studies isolate placement itself under matched hardware and routes, leaving open how vantage alone shapes downstream computer vision tasks. We address exactly this gap.

\subsection{Multi-View Sensing and Spatial Pipelines}
Navigation performance is tightly coupled to the quality of sensor data and its integration. RGB-D, stereo, and \ac{ToF} sensors provide crucial spatial depth cues\cite{multiSensor2023,tofdevice2018}, while SLAM systems fuse vision and inertial data to track pose and build maps in real time\cite{bamdad2025deep}. Devices like Meta’s Project Aria\cite{projectAria} offer tightly synchronized multi-sensor data streams optimized for such use cases.

Emerging view-synthesis methods, such as NeRF variants~\cite{gu2024egolifter} promise to work on egocentric data and in dynamic environments, suggesting that unstable cane video could be reframed as a valuable, complementary signal for 3D reconstruction rather than simply defective input. Furthermore, co-design studies consistently show that blind travelers prefer technologies that extend and augment the long cane rather than replace it~\cite{deviceNeeds2023,codesign2019,participatory_design}. 

Our work bridges this gap between XR system design and user constraints. By evaluating SLAM and NeRF performance across synchronized head and cane vantages, we establish architectural guidelines for hybrid, multi-view camera placement. We aim to determine how to leverage the stability of head-mounted tracking alongside the rich geometric detail of cane-mounted cameras, ultimately supporting more resilient assistive XR pipelines.

\subsection{Knowledge Gaps and Research Questions}

\textbf{Gap 1: Adoption and fit of emerging wearables.}
Prior surveys have examined general navigation needs of people who are blind~\cite{2019navigationsurvey,khan_technology-assisted_2018}, but rarely focus on last-mile tasks or blind cane users’ perceptions of emerging wearable placements. 
Key factors such as comfort, trust, and integration with established \ac{OandM} techniques~\cite{deviceNeeds2023,herskovitz_hacking_2023} remain underexplored, yet strongly shape adoption and are essential for interpreting technical results.

\textbf{Gap 2: Camera placement as a first-class design variable.}
Placement is usually treated as a given, yet its effect on the \emph{type and quality of spatial information} remains untested. We isolate vantage (head vs.\ cane) under controlled conditions using two diagnostic probes: \emph{(i) \ac{SLAM}} for localization and sparse mapping, and \emph{(ii) \ac{NeRF}} for dense, view-consistent 3D reconstructions; applied here for the first time to cane-mounted video.

Guided by these gaps, we ask:
\begin{itemize}[topsep=0pt]
    \item \textbf{RQ1 (Survey on adoption/fit):} How do blind cane users approach last-mile navigation today, and what are their specific spatial information needs and form-factor constraints for integrating emerging wearables (head, cane, chest) with the traditional white cane?
    \item \textbf{RQ2 (Controlled study on vantage effects):} Holding hardware and environment constant, how does camera vantage (head vs.\ cane) affect localization stability and map quality (\ac{SLAM}), and the fidelity of scene reconstructions (\ac{NeRF})? What is the specific role of cane-sweeping motion in any degradation?
    \item \textbf{RQ3 (Synthesis/design):} Given RQ1--RQ2, what design implications follow for assistive XR navigation; particularly for hybrid head+cane architectures that extend (rather than replace) the cane while minimizing cognitive load?
\end{itemize}

\noindent The survey (Section~\ref{sec:survey}) addresses RQ1; the vantage study (Section~\ref{sec:vantage}) addresses RQ2 with a stationary-cane control that separates viewpoint from sweeping motion; the Discussion (Section~\ref{sec:discussion}) synthesizes RQ3 into actionable design guidance for hybrid assistive navigation.

\section{FORMATIVE STUDY}
\label{sec:survey}



To establish assistive XR requirements, we ran a formative survey of blind cane users' last-mile strategies and their perceptions of head-, chest-, and cane-mounted wearables.

\subsection{Methods}

\paragraph{Participants}
Ten blind adults (ages 18--65+) completed the survey (\autoref{tab:participant_data}).
Eligibility required age $\geq$18, self-reported total blindness, prior \ac{OandM} training, and routine white cane use for independent travel. 
We recruited these experienced cane users via blindness community channels and word-of-mouth. All provided informed consent and received a \$30 gift card.

\begin{table*}[th!]
    \centering
    \resizebox{0.8\linewidth}{!}
    {\begin{tabular}{r r r r r r r r}
        \toprule
        ID & Age Range & Gender & Location & Years Since Blindness Onset & Years Using Cane & Travel Frequency & Cane Proficiency \\
        \midrule
        P1  & 45-54  & F & city  & since birth  & 20+     & weekly  & moderate \\
        P2  & 25-34  & M   & urban & since birth  & 11-20  & weekly  & moderate \\
        P3  & 25-34  & M   & city  & since birth  & 20+      & daily & very proficient \\
        P4  & 35-44  & F & urban & since birth  & 20+      & rarely  & very proficient \\
        P5  & 45-54  & M   & city  & 20+ years ago   & 20+  & daily & very proficient \\
        P6  & 65+    & M   & city  & since birth  & 20+    & monthly & very proficient \\
        P7  & 18-24  & M   & urban & since birth  & 6-10   & daily & very proficient \\
        P8  & 25-34  & M   & urban & since birth  & 11-20  & weekly  & very proficient \\
        P9  & 55-64  & M   & rural & 20+ years ago   & 20+    & rarely  & extremely proficient \\
        P10 & 35-44  & M  & city  & 20+ years ago  & 11-20    & daily & extremely proficient \\
        \bottomrule
    \end{tabular}}
    \caption{Demographic and mobility characteristics of survey participants ($N=10$). All participants self-identified as completely blind, reported receiving formal \ac{OandM} training, and regularly use a white cane for independent navigation. All values are self-reported.}
    \label{tab:participant_data}
\end{table*}

\paragraph{Materials \& Instrument}
The web-based survey comprised two parts. First, three co-designed scenario prompts (restaurant, crowded bus stop, shopping mall; full description in Appendix A) elicited open-ended responses regarding last-mile navigation strategies and pain points~\cite{creswell_designing_2011,lastMeter2019,unfamiliarindoorNeeds2022,2019navigationsurvey,streetnav2024,outdoorSurvey2023}. These scenarios systematically varied in acoustics, spatial structure, and landmark availability. Second, targeted Likert items probed attitudes toward navigation priorities, environmental comfort, system trust, and device placement/form factor (\autoref{fig:likert}).

\paragraph{Analysis}
Open-ended responses underwent inductive thematic analysis, with two researchers iteratively developing a shared codebook until conceptual saturation. For 5-point Likert data, we report medians and 95\% bias-corrected bootstrap confidence intervals (10,000 resamples). Given the sample size ($n{=}10$), inferential claims remain conservative; we avoid mean aggregation and instead treat individual items as reflective indicators for visualization (\autoref{fig:likert})~\cite{blandford_qualitative_2016}.

\paragraph{Ethics}
The survey protocol was reviewed by University of California, Santa Barbara Institutional Review Board (IRB) and determined to be exempt (Protocol No.\ 10-25-0112) under educational/benign behavioral research criteria. 
All procedures adhered to ethical standards for confidentiality and voluntary participation.

\subsection{Cross-Cutting Navigation Challenges and Wearable Preferences}

\begin{figure}[th!]
    \centering
    \includegraphics[width=\linewidth]{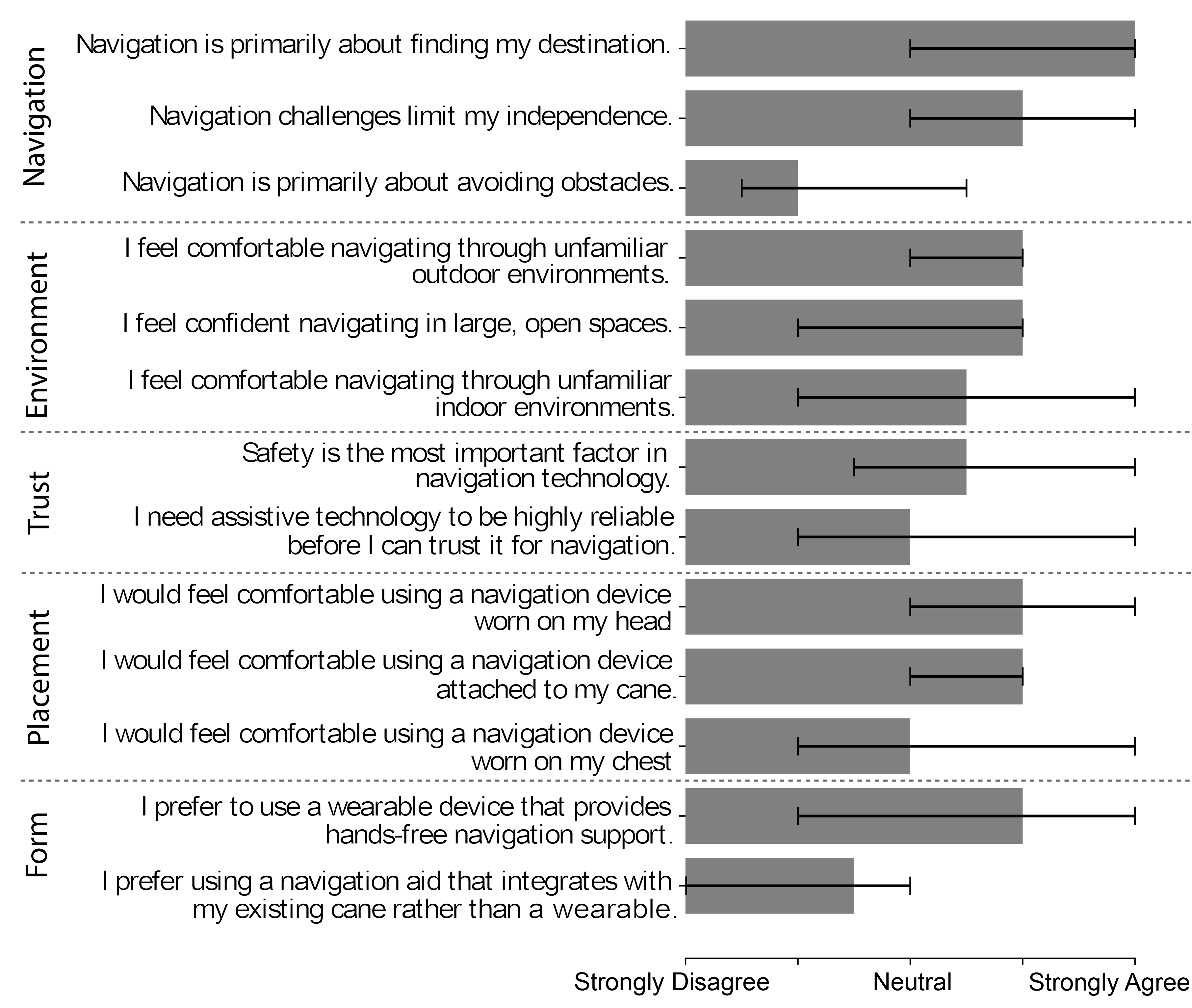}
    \caption{Participant responses to navigation and device preference questions grouped under five categories: Navigation, Environment, Trust, Placement, and Form, using a 5-point Likert scale (1 = Strongly Disagree, 2 = Somewhat Disagree, 3 = Neither Agree Nor Disagree (Neutral), 4 = Somewhat Agree, 5 = Strongly Agree). Within each category, questions are presented in descending order based on their median score. These descriptive results suggest stronger comfort with head-mounted than cane-mounted devices, alongside a preference for hands-free aids that support destination finding.}
    \label{fig:likert}
\end{figure}

We evaluated participants' prior experience with assistive technologies alongside their camera placement preferences (\autoref{fig:likert}). Participants most frequently reported prior experience with smart glasses (6), followed by smart canes (4), chest-mounted devices (3), and wrist/hand-mounted solutions (2). Because participants emphasized destination-finding over mere obstacle avoidance, head-mounted placements received the highest comfort ratings. Conversely, direct cane integration was rated poorly due to added weight, sweep interference, and social conspicuousness, highlighting a strict requirement for lightweight, unobtrusive, hands-free designs.

Beyond wearable preferences, inductive coding of open-ended responses revealed three recurring navigation challenges across varying environments:
\begin{itemize}[topsep=0pt]
    \item \textbf{Sensory overload:} High ambient noise and reverberation (e.g., in malls or at bus stops) degraded essential auditory cues and increased cognitive load, leaving participants exhausted and disoriented (P1, P4).
    \item \textbf{Crowds and social dynamics:} Dense crowds necessitated constant path adjustments. Furthermore, crowded spaces introduced social vulnerabilities (e.g., harassment concerns, P4) and unsolicited bystander assistance that often disrupted orientation rather than helping.
    \item \textbf{Infrastructure inconsistency:} Homogeneous storefront designs (P7) and the absence of tactile landmarks, such as textural changes at bus stops (P4), hindered non-visual differentiation. This lack of standardization forces inefficient, trial-and-error navigation (P9) and prevents spatial memorization (P3).
\end{itemize}

\subsection{Assistive XR Design Implications}

By coding the open-ended scenario responses and integrating them with the Likert data, we distilled four critical design constraints for assistive XR systems:

\begin{itemize}[topsep=0pt]
    \item \textbf{C1: Unobstructed Audio (Mitigating Sensory Overload).} 
        Participants described environments like malls and bus stops as acoustically chaotic. Because hearing is their primary tool beyond the cane's reach (used to track parallel traffic, echoes, and footfalls), XR audio interfaces must not occlude natural hearing. Spatial audio cues must be rendered sparsely and intelligently to prevent cognitive overload.
    
    \item \textbf{C2: Technology must complement rather than replace traditional aids.}
        The long cane is indispensable for near-field obstacle detection. Participants stressed that any cane-mounted sensors must not alter the sweeping rhythm or tactile feedback of the cane. Consequently, computer vision pipelines should not attempt to replace basic obstacle avoidance, but rather map the environment \emph{beyond} the cane's physical reach~\cite{smartcanePerception,kristjansson2016designing,kim2013usability,stanfordcane2021}.
        
    \item \textbf{C3: Trust via Robust Localization.}
        Trust emerged as a non-negotiable barrier to adoption. In environments lacking stable tactile boundaries, participants rely on trial-and-error. If an XR system provides false spatial cues due to drifting pose estimation, it actively degrades safety. Therefore, underlying SLAM pipelines must maintain rigorous, high-confidence tracking before pushing spatial audio or haptic feedback.
        
    \item \textbf{C4: The Vantage Trade-off.}
        Participants highlighted a dichotomy: head-mounted views are comfortable and stable but miss low-lying hazards and textural boundaries; cane-mounted views capture the ground plane but introduce dynamic motion that users feared would degrade system performance. 
\end{itemize}

\noindent These findings identify competing ergonomic and sensing requirements that motivate a controlled comparison of head- and cane-mounted recordings. However, the exact technical impact of placing tracking cameras on a sweeping cane which users identified as a major concern (C2 \& C4) remains empirically under-explored. To move from user perception to systems evaluation, we next isolate vantage as an independent variable to test how standard XR tracking and mapping pipelines behave under these real-world constraints.

\section{CONTROLLED CAMERA VANTAGE STUDY}
\label{sec:vantage}


Building directly on the derived XR design constraints (specifically C3 and C4), we conducted a controlled study isolating camera vantage as an independent variable. To empirically evaluate how standard XR tracking and mapping pipelines perform under real-world \ac{OandM} trajectories, we synchronized two Project Aria devices~\cite{projectAria} - mounting one on the head and one on the sweeping cane of an experienced blind traveler. By holding the hardware and environment constant across five diverse indoor and outdoor last-mile scenarios, this quantitative evaluation~\cite{creswell_designing_2011,van2014design} moves beyond user perception to measure how head versus cane-mounted perspectives impact the foundational SLAM and spatial mapping capabilities required for reliable assistive XR. We frame it as a mechanistic case study: a single expert traveler, instrumented under tightly controlled conditions, isolates the effect of camera placement on the sensing pipeline rather than estimating how that effect varies across a population. This study characterizes the quality of spatial information prior to the design of any feedback layer; it does not benchmark deployed navigation systems, nor does it evaluate audio or haptic interfaces, which depend on the upstream spatial information we measure here.

\subsection{Methods}

\subsubsection{Participant}

The participant was a blind co-author who also served as a co-designer of the study, an experienced cane user with over a decade of independent travel experience. They have no light perception and travel independently on a daily basis in urban settings using a white cane, and were formally trained in \ac{OandM}.
Their deep familiarity with non-visual travel shaped the study to reflect authentic \ac{OandM} practice. Their contributions included selecting study locations, identifying landmarks salient from a non-visual perspective, and collaborating on the interpretation of results, ensuring the evaluation remained grounded in lived experience.

\begin{figure*}[!t]
    \centering
    \includegraphics[width=0.75\textwidth]{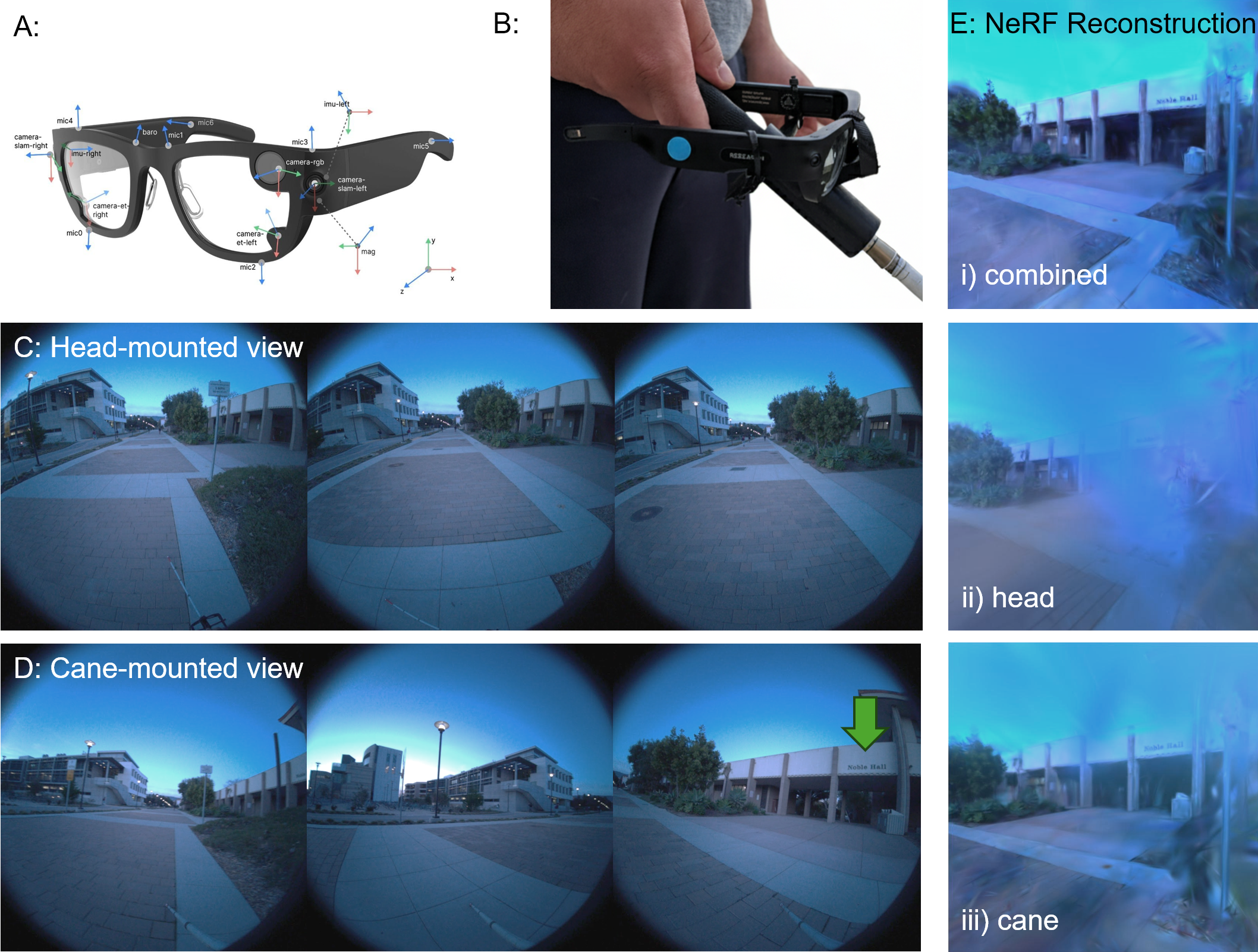}
    \caption{A) Meta's Project Aria smartglasses B) Cane Mount. The user navigates toward the entrance of Noble Hall (green arrow in D) in Outdoor2, situated next to an open plaza—a scenario where traditional canes offer limited spatial awareness. C) Head-mounted camera frames. D) Cane-mounted camera frames. E) NeRF-based 3D scene reconstructions compare head+cane (i), head-only (ii), and cane-only (iii) input streams.}
    \label{fig:vantageResults}
\end{figure*}

\subsubsection{Sensor Setup}

We used two synchronized Project Aria devices~\cite{projectAria} to record first-person video and inertial data (\autoref{fig:vantageResults}A). We selected Project Aria because it provides synchronized egocentric RGB, wide-FOV monochrome cameras, and IMU streams, along with access to the Machine Perception Services (MPS) processing pipeline. This let us compare head- and cane-mounted placement while holding both the recording hardware and the processing tools constant.
One device was mounted at the head using the standard glasses frame, while the other was attached to the cane shaft via a custom 3D-printed bracket (\autoref{fig:vantageResults}B). The cane-mounted device was strapped to the bracket with zip ties, positioned so that with the cane held at its typical 30–45° operating angle the camera sat approximately 2 feet above the ground. The bracket was angled approximately 30° relative to the cane shaft so that, at this operating angle, the camera's field of view stayed directed forward along the direction of travel.
Both devices recorded at 1408$\times$1408 resolution and 20 frames per second, with internal clocks synchronized prior to each session. 
This setup allowed for tightly synchronized video and sensor streams across both vantage points while holding hardware constant.

The cane mount was designed in consultation with an \ac{OandM} instructor to preserve natural sweep and minimize weight imbalance.
The cane, originally 300 grams, now weighs 376 grams with the attached plastic filament mount and glasses.
The blind co-author informally confirmed that the mount did not meaningfully disrupt the cane's balance or natural sweeping rhythm.

\subsubsection{Environment Selection}
Environments were chosen through a co-design process with our blind co-author, informed by the survey themes of entrance disambiguation, bus stop navigation, and large indoor orientation. 
We selected five sites to represent diverse acoustic and spatial conditions (indoor/outdoor, structured/open, static/dynamic), as described in \autoref{tab:locationStats}: two outdoor environments and three indoor environments.

\begin{figure}[tb!]
    \centering
    \includegraphics[width=\linewidth]{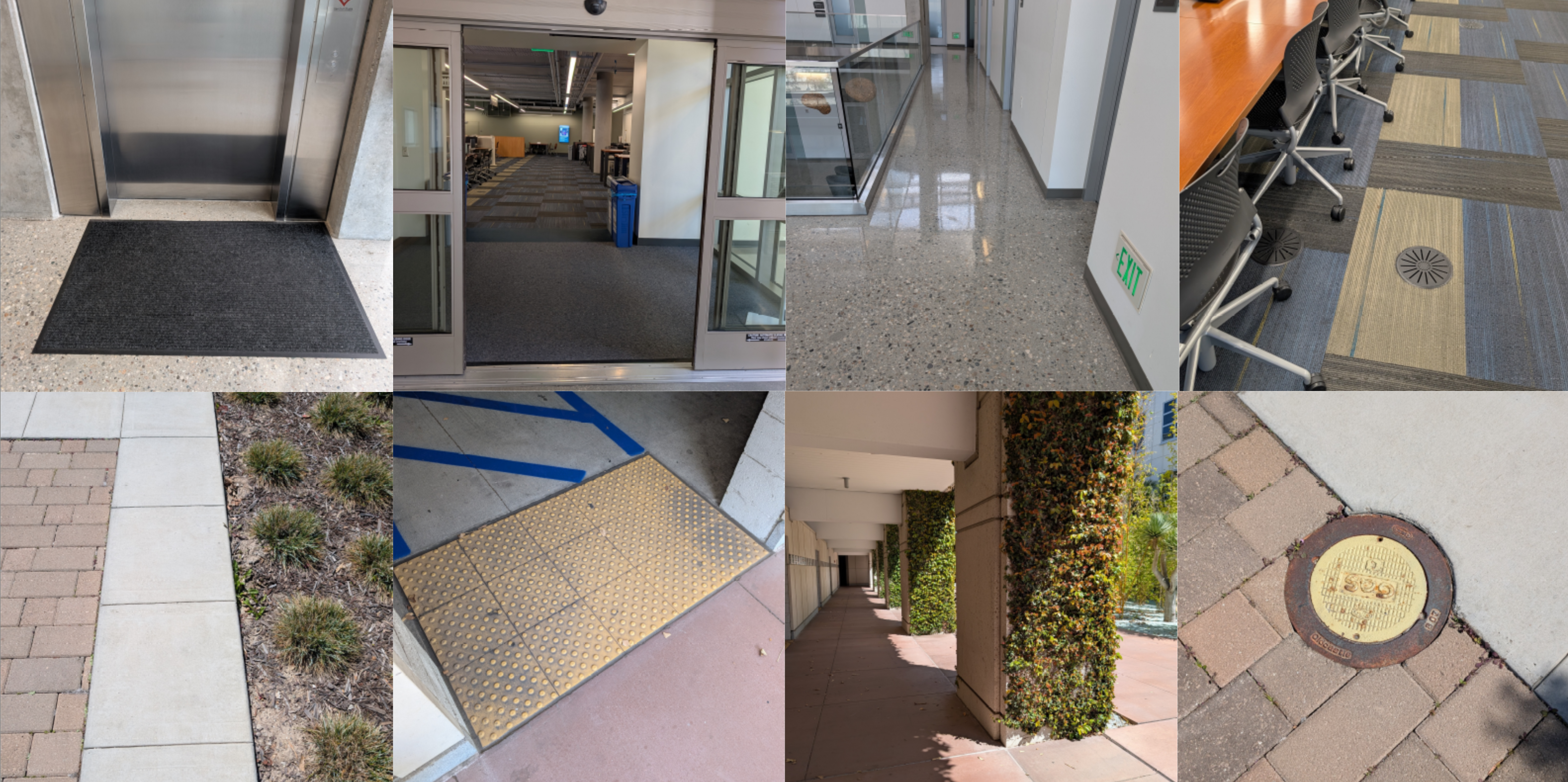}
    \caption{Examples of key visual and tactile landmarks encountered during last-mile navigation tasks, including floor transitions, textured surfaces, doorways, tactile paving, and environmental boundaries.}
    \label{fig:landmarks}
\end{figure}

Routes ranged from 25–70 meters and were designed to incorporate both tactile and visual landmarks relevant to a blind traveler (\autoref{fig:landmarks}).
Each environment presented realistic last-mile difficulties, including static obstacles (trash cans, furniture), dynamic elements (pedestrians), and variable lighting.
This process ensured ecological validity while covering the range of last-mile challenges highlighted in Section~\ref{sec:survey}.

\begin{table*}[tb!]
    \centering
    \begin{tabular}{l l m{3.6cm} p{6.2cm}}
        \toprule
        \textbf{Location} & \textbf{Distance} & \textbf{Description} & \textbf{Landmarks} \\
        \midrule
        Outdoor1 & 30m & Covered Walkway & Tactile warning pad, pillars, walls \\
        Outdoor2 & 25m & Open Sidewalk & Sidewalk-to-plant area border, manhole cover \\
        Indoor1 & 35m & Bright Modern Hallway & Floor mat at elevator, doors, railings, walls \\
        Indoor2 & 70m & Simple layout Hallway & Garbage cans, walls, floor mats at exit door \\
        Indoor3 & 45m & University Library & Garbage cans, hard-to-carpet floor transition, chairs, floor outlets, partitions \\
        \bottomrule
    \end{tabular}
    \caption{Overview of navigation locations used in the study. Each corresponds to a specific physical site with a predefined walking path (distance indicates end-to-end route length) and notable environmental landmarks.}
    \label{tab:locationStats}
\end{table*}

\subsubsection{Procedure}

At each site, the participant completed one main trial, which was designed as a loop. 
Sessions followed a standardized routine at each site: 
(i) arrival at the location; 
(ii) a brief familiarization phase with an experimenter with the cane starting from a Start landmark till a Midpoint landmark and then back to the Start landmark. All the possible intermediate landmarks along this loop were shown to the participant;
(iii) during the main trial, the participant retraced the same Start–Midpoint–Start loop independently, using standard \ac{OandM} techniques and starting from the Start landmark. Whether to make use of the intermediate landmarks shown during familiarization was left to the participant's discretion. Both devices recorded simultaneously throughout the trial;
(iv) recording was stopped once the participant completed the loop and came back to the Start landmark, marking the end of trial.

As an additional control, outside the main five-route comparison, we recorded a stationary-cane condition on one outdoor route (Outdoor2) and one indoor route (Indoor1). The participant walked the Start-to-Midpoint segment holding the cane forward and stationary rather than applying the sweeping motion. This condition shares the low cane-mounted viewpoint of normal cane use while removing the high-frequency motion sweeping introduces.

\subsubsection{Data Collection and Analysis}

Each session produced synchronized RGB video and IMU streams from both devices. 
Recordings were annotated with start and end markers, environment type, and vantage condition. 
Data were stored securely, preprocessed to ensure temporal alignment across sensors, and checked for integrity. 
Any corrupted or incomplete sessions were excluded prior to analysis.

To evaluate how vantage affects sensing performance, we analyzed the recordings using two complementary computer vision pipelines widely used in modern XR systems \cite{slam2024,sakashita2024sharednerf,kwon2023renderable}:

\noindent\textbf{\Ac{SLAM}} (via Meta's proprietary MultiSLAM framework\footnote{\url{https://facebookresearch.github.io/projectaria_tools/docs/data_formats/mps/slam/mps_multi_slam}}) was used to estimate camera poses and recover sparse 3D map points, providing a measure of localization stability and geometric anchoring. Two metrics were derived from the system’s outputs.
\begin{itemize}[topsep=0pt]
        \item \textbf{Filtered map points:} raw map points were filtered by nominal thresholds (inv\_dist\_std $\leq$ 0.005, dist\_std $\leq$ 0.01) to retain only those the pipeline reports as having low positional uncertainty. We report both the count of points passing this filter and the \emph{Map-point filter pass rate} (the proportion surviving); total raw counts additionally indicate potential coverage under more dynamic trajectories (e.g., cane sweeps).
        \item \textbf{Maximum-quality poses:} MPS assigns each camera pose a quality score in $[0,1]$. We report the count of poses assigned the maximum score of $1.0$ and the \emph{Maximum-quality pose rate} (the proportion at that score), which we use as our primary indicator of localization stability.
    \end{itemize}
We treat these metrics as diagnostic outputs rather than independent ground-truth measures of mapping accuracy: both the pose quality score and the point-uncertainty estimates are computed internally by MPS using a proprietary calculation we cannot inspect. The filtering thresholds and quality criteria we apply follow MPS's general recommendations.

\noindent\textbf{\Acp{NeRF}} were used to reconstruct view-consistent 3D scenes, offering a higher-fidelity substrate for semantic reasoning tasks such as entrance identification or free-space estimation. Specifically, we employed EgoLifter~\cite{gu2024egolifter} as a representative modern egocentric reconstruction method and diagnostic probe — chosen because it targets egocentric video with complex motion and dynamic elements, not because it was designed for cane-mounted capture. Cane mounting is a demanding tool-mounted, body-adjacent case, with stronger motion and a lower viewpoint than the standard head-mounted egocentric video these methods assume. Reconstruction fidelity was quantified using \ac{PSNR} and inspected qualitatively for coverage and coherence. For the combined Head+Cane condition, frames from both devices were supplied jointly as multi-view input to a single EgoLifter reconstruction; we did not perform any explicit runtime pose fusion between the two streams.

This dual analysis allowed us to examine how camera vantage affects two core spatial computing functions, localization and scene reconstruction, that downstream assistive XR systems may depend on.

\subsubsection{Ethics}
The vantage study involved data collection with a blind traveler in real environments and underwent full review and approval by University of California, Santa Barbara IRB (Protocol No.\ 9-24-0470). 

The participant provided informed consent. 
Recordings were anonymized using EgoBlur \cite{raina2023egoblur} before analysis, and all procedures prioritized the participant's safety, comfort, and agency throughout the study.

\subsection{Results}

We evaluated vantage effects on \ac{SLAM} performance and \ac{NeRF}-based 3D scene reconstruction. 
Results are summarized in \autoref{tab:slam_performance}, \autoref{tab:slamAblation}, and \autoref{tab:nerfResults}, with qualitative examples shown in \autoref{fig:vantageResults} (see Appendix C for more). 
Below we highlight key findings.

\subsubsection{SLAM Performance}

As summarized in \autoref{tab:slam_performance}, the head-mounted camera consistently yielded higher values on both SLAM metrics. Its Maximum-quality pose rate exceeded 98\% across all sites, while the cane-mounted rate dropped to 55–58\% in outdoor settings. 
Although cane sweeps generated more raw map points overall, only 19–27\% passed the uncertainty filter, compared to 27–39\% for the head-mounted camera. 
These results suggest a consistent trade-off across the evaluated routes: head placement supports more stable localization, whereas cane placement captures broader coverage but with reduced reliability.

\begin{table*}[!tb]
\centering
\resizebox{0.9\linewidth}{!}{\begin{tabular}{lrrrr}
\toprule
\textbf{Location} & \multicolumn{2}{c}{\textbf{Filtered map points $\uparrow$}} & \multicolumn{2}{c}{\textbf{Maximum-quality poses $\uparrow$}} \\
\cmidrule(lr){2-3} \cmidrule(lr){4-5}
& \multicolumn{1}{c}{\textbf{Cane}} & \multicolumn{1}{c}{\textbf{Head}} & \multicolumn{1}{c}{\textbf{Cane}} & \multicolumn{1}{c}{\textbf{Head}} \\
\midrule
Outdoor1 & 26.5\% ({\small 361,373 / \textbf{1,364,893}}) & \textbf{35.9\%} ({\small \textbf{390,267} / 1,087,275}) & 58.2\% ({\small 69,125 / \textbf{118,771}}) & \textbf{98.3\%} ({\small \textbf{111,789} / 113,676}) \\
Outdoor2 & 27.3\% ({\small 81,252 / \textbf{297,735}}) & \textbf{34.5\%} ({\small \textbf{99,566} / 288,453}) & 55.1\% ({\small 20,010 / 36,300}) & \textbf{99.9\%} ({\small \textbf{42,759} / \textbf{42,810}}) \\
Indoor1 & 27.3\% ({\small \textbf{449,647} / \textbf{1,645,114}}) & \textbf{39.7\%} ({\small 442,985 / 1,116,486}) & 84.7\% ({\small 166,447 / 196,584}) & \textbf{99.9\%} ({\small \textbf{196,912} / \textbf{197,014}}) \\
Indoor2 & 19.6\% ({\small \textbf{297,709} / \textbf{1,520,318}}) & \textbf{27.4\%} ({\small 281,707 / 1,029,381}) & 73.2\% ({\small 150,316 / \textbf{205,414}}) & \textbf{99.4\%} ({\small \textbf{200,814} / 201,988}) \\
Indoor3 & 22.4\% ({\small \textbf{498,057} / \textbf{2,223,797}}) & \textbf{34.1\%} ({\small 471,954 / 1,382,517}) & 94.6\% ({\small 201,190 / \textbf{212,647}}) & \textbf{99.9\%} ({\small \textbf{207,038} / 207,089}) \\
\bottomrule
\end{tabular}}
\caption{Comparison of SLAM performance across five navigation environments, showing the percentage (pass rate) and raw counts of Filtered map points and Maximum-quality poses ($\uparrow$ higher is better), reported as a fraction of total map points or pose estimates. Filtered map points indicate potential coverage, and the Maximum-quality pose rate indicates localization stability. Bolded values indicate the best-performing configuration within each environment.}
\label{tab:slam_performance}
\end{table*}

\subsubsection{Stationary-Cane Control}
To isolate the effect of motion, we compared three configurations: head-mounted, 
cane-mounted (dynamic sweep), and cane-mounted (stationary ``no sweep'').
Holding the cane steady improved the outdoor Maximum-quality pose rate from 39.5\% to 83.2\%, 
substantially narrowing the gap with the head-mounted rate (99.6\%, 
\autoref{tab:slamAblation}). 
This suggests that the reduced pose quality was associated primarily with sweeping motion rather than the low viewpoint itself.

\begin{table*}[!t]
\centering
\resizebox{0.85\linewidth}{!}{\begin{tabular}{llrrr}
\toprule
\textbf{Location} & \textbf{Performance Metric} & \textbf{Head-Mounted} & \textbf{Cane-Mounted} & \textbf{Cane-Mounted} \\
& & \textbf{Camera} & \textbf{Camera} & \textbf{(No Sweep)} \\
\midrule
\multirow[t]{2}{*}{Outdoor} & Filtered map points $\uparrow$ & \textbf{37.6\%} {\small (\textbf{50,531} / 135,285)} & 26.6\% {\small (40,846 / \textbf{153,566})} & 30.1\% {\small (44,744 / 148,832)}\\
    & Maximum-quality poses $\uparrow$ & \textbf{99.6\%} {\small (\textbf{13,315} / 13,366)} & 39.5\% {\small (6,868 / \textbf{17,408})} & 83.2\% {\small (11,095 / 13,338)} \\
\cdashline{1-5}
\multirow[t]{2}{*}{Indoor} & Filtered map points $\uparrow$ & 38.0\% {\small (82,448 / 217,227)} & 34.2\% {\small (101,556 / \textbf{296,521})} & \textbf{39.2\%} {\small (\textbf{115,779} / 295,671)} \\
    & Maximum-quality poses $\uparrow$ & \textbf{99.8\%} {\small (25,561 / 25,613)} & 92.6\% {\small (\textbf{27,437} / \textbf{29,628})} & 98.5\% {\small (26,167 / 26,575)} \\
\bottomrule
\end{tabular}}
\caption{Comparison of SLAM performance across indoor and outdoor environments, evaluating standard head- and cane-mounted configurations against a cane-mounted condition with sweeping motion disabled (``No Sweep"). Metrics include the pass rate (percentage) and raw counts of Filtered map points and Maximum-quality poses ($\uparrow$ higher is better), reported as a fraction of total points or pose estimates. Filtered map points are those passing the MPS uncertainty thresholds; Maximum-quality poses are those assigned the maximum MPS quality score (1.0). Bolded percentages indicate the best-performing configuration for each metric within each environment.}
\label{tab:slamAblation}
\end{table*}

\subsubsection{NeRF-Based 3D Scene Reconstruction}

Reconstruction quality differed significantly by configuration (\autoref{tab:nerfResults}). 
A hierarchical mixed-effects model, with per-frame PSNR of held-out views as the unit of analysis and Scenario (indoor vs. outdoor) and Location (nested within Scenario) as random effects, showed a significant fixed effect of configuration on PSNR. \footnote{Formula: \texttt{PSNR \textasciitilde{} Configuration + (1 | Scenario/Location)}} This structure captures variability at multiple levels of the environment hierarchy and isolates the effect of camera placement.
Combined head+cane input achieved the highest PSNR, outperforming cane-only (Estimate = 1.204, SE = 0.193, $p < .001$) and head-only (Estimate = 1.811, SE = 0.194, $p < .001$). 
Cane-only also outperformed head-only (Estimate = 0.607, SE = 0.223, $p = .007$), suggesting its ground-level perspective adds complementary cues. 
Random effects indicated variability across Scenarios ($\sigma^2 = 7.346$) and Locations ($\sigma^2 = 5.906$), with PSNR generally lower outdoors, reflecting greater complexity and inconsistent lighting. 
Qualitative reconstructions (\autoref{fig:vantageResults} and Appendix C) likewise showed that hybrid input produced more coherent geometry, especially in cluttered indoor and sparse outdoor settings.

\begin{table}[!t]
    \begin{tabular}{lccc}
    \toprule
    \textbf{Location} & \textbf{Cane} & \textbf{Head} & \textbf{Head+Cane} \\
    \midrule
    Outdoor1 & 12.95 ± 3.34 & 13.32 ± 3.85 & \textbf{13.89 ± 3.91} \\
    Outdoor2 & 12.66 ± 1.75 & 16.23 ± 3.35 & \textbf{17.87 ± 2.82} \\
    Indoor1 & 19.56 ± 2.41 & 19.73 ± 2.00 & \textbf{22.21 ± 2.74} \\
    Indoor2 & 21.25 ± 2.18 & 18.44 ± 2.13 & \textbf{21.50 ± 2.46} \\
    Indoor3 & 16.22 ± 2.33 & 15.76 ± 2.55 & \textbf{16.46 ± 2.19} \\
    \cdashline{1-4}
    Outdoor & 12.89 ± 3.04 & 14.12 ± 3.92 & \textbf{14.90 ± 4.05} \\
    Indoor & 18.96 ± 3.13 & 17.94 ± 2.78 & \textbf{19.99 ± 3.57} \\
    \bottomrule
    \end{tabular}
    \caption{The table reports PSNR ($\uparrow$ higher is better)  for each location under three input configurations: Cane-mounted, Head-mounted, and Head+Cane. Bold values indicate the highest mean PSNR in each row.}
    \label{tab:nerfResults}
\end{table}

\subsection{Summary of Findings}

Together, these empirical results highlight a trade-off in assistive XR design: the tension between tracking stability and near-field spatial coverage. Head-mounted sensors yielded a higher Maximum-quality pose rate and a higher filter pass rate for map points, indicating the head as the more stable vantage in our evaluated pipelines for robust inside-out SLAM (addressing Constraint C3). Conversely, cane-mounted cameras suffered substantial tracking degradation, consistent with the erratic trajectories of the \ac{OandM} sweep. A stationary-cane control indicated that this localization failure is associated primarily with sweeping motion rather than the low viewpoint itself (Constraint C2). However, despite poor tracking, the cane vantage successfully captured critical near-field, ground-level geometry largely missed by the head.

Crucially, the hybrid head-plus-cane configuration achieved the highest overall 3D reconstruction fidelity. This suggests that a hybrid architecture — anchoring \ac{SLAM} pose estimation on the stable head while mapping the ground plane via the cane — is a promising direction for assistive spatial computing, though we evaluate it here only as a combined multi-view input to reconstruction rather than as a real-time fusion system. By grounding user perceptions (Constraint C4) in measured sensing pipeline performance, we show that head and cane vantages are empirically complementary in our setting rather than interchangeable. 


\section{GENERAL DISCUSSION}
\label{sec:discussion}

This work examines camera placement as a central design variable in wearable navigation aids for blind cane users. Integrating results across our two studies, we find that survey perceptions of cane instability were corroborated by the lower measured \ac{SLAM} metrics, while survey skepticism about cane integration was challenged by the cane stream's complementary reconstruction benefits — a convergence that motivates the hybrid approaches discussed below.


\subsection{Camera Placement Shapes Perception and Performance}

Camera placement is not a one-size-fits-all decision but must be tailored to task and environment. 
Survey participants rated head-, chest-, and cane-mounted devices as somewhat acceptable (\autoref{fig:likert}), with preferences shaped by familiarity: head-mounted devices such as smartglasses were widely recognized, while skepticism toward cane-mounted systems reflected prior underperformance~\cite{caneLessons2020,stanfordcane2021,khan_technology-assisted_2018}. 

Quantitative evaluation underscored a trade-off between localization stability and spatial coverage. 
Head-mounted cameras consistently yielded higher values on \ac{SLAM} metrics, aligning with prior work~\cite{motioBlur2021}, while cane-mounted views captured unique low-level details~\cite{arthur_effects_2000,dlbaseddevice2019}, evident in \ac{NeRF}-based reconstructions (\autoref{fig:vantageResults}). 
The stationary-cane control suggests that sweeping motion was a major contributor to reduced MPS pose quality on the two evaluated partial routes (\autoref{tab:slamAblation}).
Rather than discarding cane-mounted vision, future systems should adapt to its dynamics with stabilization or motion-aware algorithms. Advancing tracking capabilities for this use case could benefit the broader XR community by making more robust inside-out tracking for headsets (e.g., VR sports games also have a lot of high frequency motion and need reliable tracking).

\subsection{Implications for Assistive Technology Design}

From a design perspective, our findings emphasize that navigation technologies must work in harmony with established practices. 
Participants emphasized the irreplaceable role of tactile cane feedback for near-field obstacle detection~\cite{williams_just_2014}, noting that its utility cannot be replicated by other modalities. 
At the same time, tasks such as bus identification or navigating large open indoor spaces extend beyond the cane's reach. 
Participants expressed a preference for hands-free solutions that complement, not override, existing cane use. 
Head-mounted sensing offers stability and ease of adoption, while cane-mounted sensing contributes complementary ground-level information that improves scene reconstruction when supplied jointly with head-level input. 
This openness to cane-mounted sensing, despite mixed prior experiences, underscores a crucial design principle: technologies that align with established \ac{OandM} practice are more likely to be trusted and adopted, while tools that interfere with sweep or attempt to replace tactile feedback risk rejection and cognitive overload. 
Future assistive systems should therefore adopt adaptive or hybrid vantage strategies that explicitly account for cane feedback, balancing localization stability with broader environmental coverage. 
More broadly, the need to combine stable and dynamically moving viewpoints is relevant to XR systems that rely on distributed body-worn sensors.

\subsection{Toward Adaptive, Multisensor Navigation Systems}

Because we measure the sensing layer rather than an interface, our clearest implications are constraints on what a feedback layer can draw upon, supporting tasks such as entrance identification, free-space estimation, and path planning. The head-mounted stream's stable localization is the more trustworthy basis for \emph{global} cues such as heading, orientation, and route-level guidance, where drift would directly mislead the traveler (Constraint~C3), while cane-derived geometry, though poor for localization, uniquely captures the near-ground structure the head misses, making it the better source for \emph{local}, selective alerts surfaced only when relevant. This mirrors strategies cane users already employ, drawing simultaneously on tactile, auditory, and spatial-memory cues.

This division also governs how information reaches the traveler without overloading it: because the physical cane already conveys ground textures and curbs by touch, the \ac{XR} system need not re-render these via audio, and can reserve spatial audio for objects beyond the cane's reach, such as overhanging branches, approaching vehicles, or specific storefronts. This addresses the fragility of auditory reliance our survey identified (Constraint~C1), and the same principles of modality offloading and selective rendering apply to \ac{AR} for sighted users, whose audio or haptic cues must not compete with visual attention on the physical world.

Realizing this will require advances in motion compensation, lightweight multi-view integration, and context-aware feedback, together with co-design to ensure such systems reduce cognitive burden and integrate with established \ac{OandM} technique.

\subsection{Limitations and Future Work}
\label{sec:limitations}

Despite these contributions, several limitations point to opportunities for future work.

The controlled vantage study involved a single blind participant and should be read as a mechanistic case study of camera placement under authentic cane use, not as evidence for population-level differences across blind travelers. This design let us hold participant behavior, hardware, routes, and environments constant while comparing synchronized head- and cane-mounted streams across five real-world locations, isolating the sensing question in a way a multi-user study cannot at this level of control. It is therefore appropriate for characterizing how vantage shapes the signal available to \ac{SLAM} and \ac{NeRF} pipelines, but it does not establish how those effects vary across travelers, cane techniques, or gaits, and does not replace a broader user study.

Second, our study environments were intentionally bounded. The survey used three analytic scenarios (restaurant, bus stop, mall) and the vantage study examined five real-world routes, chosen for clarity and comparability. While they captured variation in spatial and sensory demands, they do not reflect the full diversity of blind travel; future work should extend evaluation to dynamic, crowded settings that surface the broader social and sensory complexities raised in the survey~\cite{blandford_qualitative_2016}.

Third, we evaluated vantage using \ac{SLAM} and \ac{NeRF} as diagnostic probes rather than integrated real-time systems. This allowed us to isolate placement effects, but stops short of demonstrating end-to-end deployment. Follow-up work should validate these trends in real-time assistive pipelines, advancing lightweight \ac{SLAM}, multi-view reconstruction, and camera stabilization.

Finally, differences in scale between the survey (multi-user) and vantage study (single-user) limit direct comparability. While this constrains integration validity, triangulating across methods strengthens confidence in the core finding: head- and cane-mounted perspectives are complementary.

\section{CONCLUSION}

This work shows that camera placement is as critical to assistive navigation design as the data processing techniques and algorithms. By combining community-grounded perspectives with a controlled evaluation of vantage, our results indicate that head-mounted cameras provide stable localization while cane-mounted cameras contribute complementary ground-level detail; together offering richer, more reliable scene understanding than either alone.

These findings underscore that future assistive XR systems should augment, not replace, the tactile foundation of the cane, integrating head- and cane-mounted perspectives to balance stability with coverage. Hybrid approaches that respect established \ac{OandM} practices hold promise for enabling confident, independent last-mile travel.

Looking ahead, large-scale, multi-user datasets collected in diverse real-world conditions will be essential to refine these hybrid designs, enabling spatial AI systems that fuse multiple viewpoints, adapt to the dynamics of cane use, and earn the trust of blind travelers.
By designing for the demanding constraints of blind pedestrians, this work points towards assistive XR systems that better balance localization stability with near-ground scene coverage, while also suggesting broader lessons for robust embodied sensing in XR.



\acknowledgments{
We thank Brianna Pettit for giving insights into \ac{OandM} instruction and our undergrad Research Assistants for helping fabricate cane mount for the glasses.
Supported by the National Library of Medicine of the National
Institutes of Health (NIH) under Award Number DP2-LM014268.
The content is solely the responsibility of the authors and does not
necessarily represent the official views of the NIH.
In accordance with IEEE guidelines, generative AI tools were used in limited ways during manuscript preparation. Gemini 3 Pro (Google) was used for light editing and language refinement of portions of the manuscript. All AI-generated outputs were reviewed and verified by the authors, who take full responsibility for the content of the article.}

\section*{Supplemental Materials}
\label{sec:supplemental_materials}

Supplemental materials include the full survey instrument, additional qualitative results, SLAM point clouds and NeRF reconstructions for all five locations, and documentation of the co-design process. A video shows synchronized head- and cane-mounted recordings from one representative route. We also release the full dataset and derived analysis outputs under a CC BY 4.0 license at \url{https://huggingface.co/datasets/apurv4166/LastMileEvaluation}.

\bibliographystyle{abbrv-doi}

\bibliography{references}
\end{document}